\documentclass[preprint2]{aastex}

\slugcomment{submitted to ApJ}

\shorttitle{X-Rays from the Jets of XTE J1550-564}
\shortauthors{Kaaret et al.}

\begin{document}

\title{X-Ray Emission from the Jets of XTE J1550--564}

\author{P.\ Kaaret\altaffilmark{1}, S.\ Corbel\altaffilmark{2}, J.A.\
Tomsick\altaffilmark{3}, R.\ Fender\altaffilmark{4}, J.M.\
Miller\altaffilmark{5}, J.A.\ Orosz\altaffilmark{6}, A.K.\
Tzioumis\altaffilmark{7}, R.\ Wijnands\altaffilmark{5}}

\altaffiltext{1}{Harvard-Smithsonian Center for Astrophysics, 60 Garden
St., Cambridge, MA 02138, USA; pkaaret@cfa.harvard.edu}

\altaffiltext{2}{Universit\'e Paris VII and Service d'Astrophysique,
CEA, CE-Saclay. 91191 Gif sur Yvette, France}

\altaffiltext{3}{Center for Astrophysics and Space Sciences, University
of California at San Diego, La Jolla, CA 92093-0424, USA}

\altaffiltext{4}{Astronomical Institute `Anton Pannekoek', University
of Amsterdam and Center for High Energy Astrophysics, Kruislaan 403,
1098 SJ Amsterdam, The Netherlands}

\altaffiltext{5}{Center for Space Research, Massachusetts Institute of
Technology, 77 Massachusetts Avenue, Cambridge, MA 02139, USA; Chandra
Fellow}

\altaffiltext{6}{Astronomical Institute, Utrecht University, Postbus
80000, 3508 TA Utrecht, The Netherlands}

\altaffiltext{7}{Australia Telescope National Facility, CSIRO, P.O.\
Box 76, Epping, NSW 1710, Australia}

\begin{abstract}

We report on X-ray observations of the the large-scale jets recently
discovered in the radio and detected in X-rays from the black hole
candidate X-ray transient and microquasar XTE J1550--564.  On 11 March
2002, X-ray emission was detected $23\arcsec$ to the West of the black
hole candidate and was extended along the jet axis with a full width at
half maximum of $1.2\arcsec$ and a full width at 10\% of maximum
intensity of $5\arcsec$.  The morphology of the X-ray emission matched
well to that of the radio emission at the same epoch.  The jet moved by
$0.52\arcsec \pm 0.13\arcsec$ between 11 March and 19 June 2002.  The
apparent speed during that interval was $5.2 \pm 1.3 \rm \, mas/day$. 
This is significantly less than the average apparent speed of $18.1 \pm
0.4 \rm \, mas/day$ from 1998 to 2002, assuming that the jet was
ejected in September 1998, and indicates that the jet has decelerated. 
The X-ray spectrum is adequately described by a powerlaw with a photon
index near 1.8 subject to interstellar absorption.  The unabsorbed
X-ray flux was $3.4 \times 10^{-13} \rm \, erg \, cm^{-2} \, s^{-1}$ in
the 0.3-8~keV band in March 2002, and decreased to $2.9 \times 10^{-13}
\rm \, erg \, cm^{-2} \, s^{-1}$ in June.  We also detect X-rays from
the eastern jet in March 2002 and show that it has decelerated and
dimmed since the previous detections in 2000.  

\end{abstract}


\section{Introduction}

Jets are a ubiquitous facet of accretion in systems ranging from young
stellar objects, to Galactic X-ray binaries, to active galactic
nuclei.   Many important questions regarding jet formation and
propagation remain unanswered and new data will likely be required to
resolve them.  Observations of systems within the Galaxy hold the great
advantage that their evolution is rapid.  Processes requiring millions
of years in AGN can unfold in a few years for stellar-mass scale
systems, so that the dynamics can be studied directly.

\citet{corbel_iau,corbel_jet} recently discovered a large-scale,
relativistic radio and X-ray emitting jet from the X-ray transient XTE
J1550--564.  The source is a Galactic black hole candidate and the mass
of the compact object is constrained to be $8 - 12 M_{\odot}$
\citep{orosz02}.  A superluminal jet ejection event had been previously
observed from this source in the radio in September 1998
\citep{hannikainen01} during a very bright X-ray outburst
\citep{sobczak99}.  Detection of the large-scale radio jet to the west
of XTE J1550--564 led to a re-analysis of archival Chandra data and
discovery of an X-ray jet to the east of XTE J1550--564
\citep{corbel_jet,tomsick_jet}.  This X-ray jet exhibits proper motion
directly away from XTE J1550--564.  Based on the angular separations
and the measured proper motion, the most likely epoch of origin of
these large-scale jets is near the major X-ray outburst which peaked on
20 Sept 1998 \citep{sobczak99} and we adopt this date as the origin of
the currently observed jets \citep{corbel_jet}.

Discovery of the large-scale radio jet also led to new observations of
XTE J1550--564 made with the Chandra X-Ray  Observatory
\citep{chandra_ref} which are described here. The sub-arcsecond
resolution of Chandra delivered by the High-Resolution Mirror Assembly
\citep{hrma_ref} has allowed us to detect and resolve X-ray emission
from the western jet.  In addition, we have again detected the eastern
jet.  In the following, \S 2 presents the new observation and source
detections, \S 3 contains our results on the spectrum, morphology,
motion, and variability of the eastern jet, \S 4 describes our new data
on the western jet, and \S 5 includes a few comments on the
implications of these results.

% \epsscale{1.8} for preprint2 1.0 for preprint

\begin{figure*} \centerline{\epsscale{1.8} \plotone{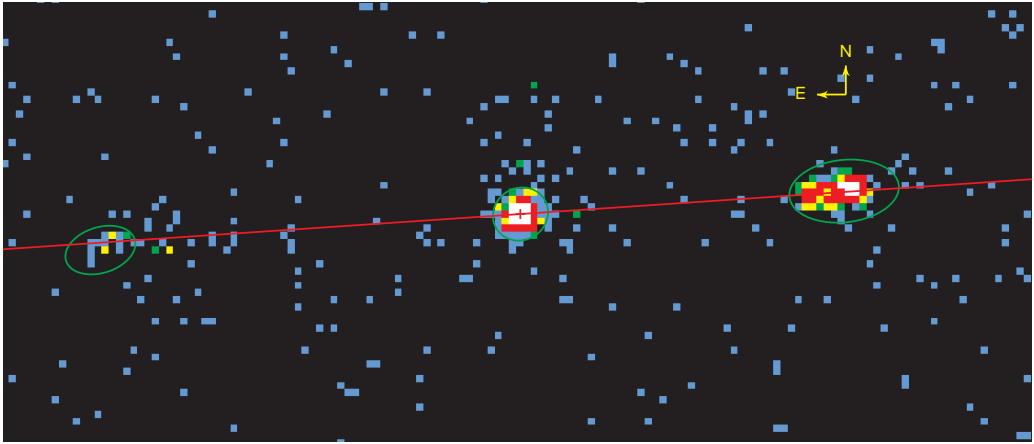}}
\caption{X-ray image of XTE J1550--564 for the 0.3--7~keV band taken on
11 March 2002.  The color of each pixel represents the number of X-ray
counts: black = 0 counts, blue = 1, green = 2, yellow = 3--4, red =
5-20, white = 21-330.  The green ellipses are source regions and
indicate detection of the western and eastern jets in addition to XTE
J1550--564 itself.  The red line is the axis of the superluminal jet
emission at a position angle of $-86.1\arcdeg$ (D.\ Hannikainen,
private communication).  The arrows indicating north and east are
$2\arcsec$ long.  \label{ximage}} \end{figure*}

\begin{deluxetable}{llccrrrl}
%\tabletypesize{\scriptsize}
\tablecaption{X-ray sources near XTE J1550--564
  \label{tablesrc}}
\tablewidth{0pt}
\tablehead{ & Date &
  \colhead{RA} & \colhead{DEC} & 
  \colhead{S/N} & 
  \colhead{Counts} & 
  \colhead{Flux}  &
  \colhead{Comment} }
\startdata
1 & March & 15 50 58.66 & -56 28 35.2 & 334.3 & 1163.4 &  4.64 & XTE J1550--564 \\
  & June  & 15 50 58.66 & -56 28 35.2 &  26.3 &   54.5 &  0.31 \\
2 & March & 15 50 55.97 & -56 28 33.6 & 122.5 &  409.1 &  1.64 & Western Jet\\
  & June  & 15 50 55.90 & -56 28 33.6 &  87.4 &  238.2 &  1.36 \\
3 & March & 15 51 02.16 & -56 28 37.7 &   4.9 &   18.4 &  0.08 & Eastern Jet \\
  & June  &             &             &       &        &       & Not detected \\
\enddata

\tablecomments{Table~\ref{tablesrc} includes for each source: RA and
DEC -- the position of the source in J2000 coordinates; S/N -- the
significance of the source detection as calculated by {\it wavdetect};
Counts - the net counts in the 0.3--7~keV band; Flux -- the observed
(absorbed) source flux in units of $10^{-13} \, \rm erg \, cm^{-2} \,
s^{-1}$ in the 0.3--7~keV band calculated assuming a power law spectrum
with photon index of 1.8 and Galactic absorption.  There are two rows
for each source: one for the March observation and one for the June
observation.} \end{deluxetable}

\section{Observations and Source Detection}

XTE J1550--564 was observed on 11 March 2002 and on 19 June 2002 with
the Chandra X-Ray Observatory (CXO; Weisskopf et al.\ 2002) using the
the Advanced CCD Imaging Spectrometer spectroscopy array (ACIS-S; Bautz
et al.\ 1998) using Director's Discretionary Time in response to a
Target of Opportunity request based on the discovery of the western
radio jet \citep{corbel_iau} and a follow-up request made after
discovery of X-ray emission from the western jet.  For the March
observation (ObsID 3448), a total of 26118~s of useful exposure were
obtained with the ACIS-S operated in a 1/4 subarray mode with only the
S3 chip read out to minimize pile-up for XTE J1550--564 which was in
the decay phase of a recent X-ray outburst and expected to be
relatively bright during the observation.  The roll angle of
$68\arcdeg$ placed the short axis of the field of view in the subarray
mode nearly along the jet axis, but the useful field of view along that
axis, $\sim 120\arcsec$, was still adequate to image both jets.  For
the June observation (ObsID 3672), a total exposure of 18025~s was
obtained with the ACIS-S operated in the full array mode with 6 ACIS
chips read out.

The data were subjected to the standard data processing (ASCDS version
6.6.0 using CALDB version 2.12) and then customary event processing and
filtering procedures, from the Chandra Interactive Analysis of
Observations software package ({\it CIAO}) v2.2.1, were applied to
produce a level-2 event list.  Light curves including all valid events
on the S3 chip were constructed to search for times of high
background.  In both observations, the count rate appears uniform with
no obvious flares.

Images of the region around XTE J1550--564 extended $120\arcsec$ along
the jet axis and $60\arcsec$ perpendicular to the jet axis were
constructed using photons in the energy band 0.3--7~keV. We searched
for sources using {\it wavdetect} \citep{freeman02}, the wavelet-based
source detection routine in {\it CIAO}.  Four sources were detected at
significances greater than $4\sigma$.  One source is coincident with
XTE J1550--564 and two others lie along the jet axis.  The western jet
was detected in both March and June, while the eastern jet was detected
only in March.  The fourth source, at RA = 15h50m53$\arcsec$.05, DEC =
-56$\arcdeg$29$\arcmin$02$\arcsec$.0 (J2000), does not lie along the
jet axis and is identified with a relatively nearby K/M dwarf present
in optical images of the field.  A portion of the March image along the
jet axis and containing XTE J1550--564 and its two jets is shown in
Fig.~\ref{ximage}.

We adjusted the astrometry of each image so that the position of the
source coincident with XTE J1550--564 matches the accurately measured
radio position.  A shift of $0.76\arcsec$ was required for the March
observation and a shift of $0.22\arcsec$ for June.  Both shifts are
within the absolute astrometric accuracy of Chandra. The statistical
error on the source positions is less than $0.22\arcsec$ ($1 \sigma$)
in all cases and less than $0.1\arcsec$ for XTE J1550--564.  However,
care must be taken in comparing positions of the jet sources due to
their finite extent as discussed below.

We defined source regions using ellipses with radii 4 times the radius
of 50\% encircled energy calculated from the observed photons for each
source by {\it wavdetect}.  The diameter of the ellipse for XTE
J1550--564 is $3.9\arcsec$ which is consistent with that expected for
an on-axis point source (8 times the $\sim 0.5\arcsec$ half power
radius for ACIS-I). We extracted counts for each source region and a
corresponding background region.  The exposure was calculated for a
monochromatic beam of 1.5~keV.  The exposure over the regions including
XTE J1550--564 and the western jet is uniform to within 2\%.  The
region including the eastern jet lies over a CCD node boundary and the
exposure is up to 15\% lower.  We translated the count rates to photon
fluxes using the exposure, then to energy fluxes in the 0.3-7~keV band
assuming a powerlaw spectrum with a photon index of 1.8 and a column
density equal to the total Galactic column in the direction of XTE
J1550--564 which we take to be $N_H = 9.0 \times 10^{21} \rm \,
cm^{-2}$ based on radio measurements of the H{\sc i} column density
\citep{dickey90} along the line of sight.  This $N_H$ value is
consistent with the one measured by high-resolution X-ray spectroscopy
of XTE J1550--564 by \citet{miller02}.  The properties of the three
sources along the jet axis are listed in Table~\ref{tablesrc}.

Below, we also include results from archival Chandra data.  
Observations with two dimensional imaging are available for June 9,
August 21, and September 11 of 2000.  These data and the analysis of
the eastern jet are described in \citet{tomsick_jet}.  Each observation
is brief with an exposure of no more than 5200~s.  The June 2000
observation had a grating in place, while the others did not.

\begin{figure}[t] \centerline{\epsscale{0.7} \plotone{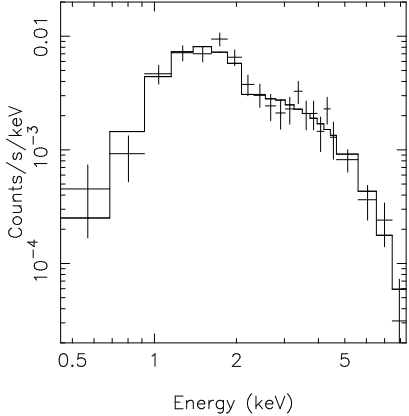}}
\caption{X-ray spectrum of emission from the western jet in the March
observation.  The curve is the best fit powerlaw model with absorption
fixed to the Galactic value. \label{srcr_spec}}   \end{figure}

\section{Western Jet}

\subsection{Spectrum}

We extracted a spectrum for the western jet from the March observation
using {\it CIAO} tools and applied the recent correction for the low
energy quantum efficiency degradation of the ACIS.  We fitted the
spectrum using XSPEC, see Fig.~\ref{srcr_spec}.  The spectrum is
adequately fit with a powerlaw model with absorption by material with
solar abundances.  With the equivalent Hydrogen absorption column
density fixed to $N_{\rm H} = 9.0 \times 10^{21} \rm \, cm^{-2}$, the
best fit photon index is $\Gamma = 1.77$ and the allowed range at 90\%
confidence is 1.61 to 1.93 ($\Delta \chi^2 = 2.7$ for one parameter). 
The absorbed flux is $(1.9 \pm 0.1) \times 10^{-13} \rm \, erg \,
cm^{-2} \, s^{-1}$ in the 0.3-8 keV band and unabsorbed flux would then
be $3.4 \times 10^{-13} \rm \, erg \, cm^{-2} \, s^{-1}$ in the same
band.

Allowing the absorption column density to vary leads to an allowed
range of $N_{\rm H} = (7.9-14.7) \times 10^{21} \rm \, cm^{-2}$ at 90\%
confidence ($\Delta \chi^2 = 4.6$ for two parameters) and a
corresponding range of 1.58--2.37 for the photon index.  The best fit
values are $N_{\rm H} = 10.9 \times 10^{21} \rm \, cm^{-2}$ and $\Gamma
= 2.00$.  The allowed $N_{\rm H}$ range includes the Galactic H{\sc i}
value.  We cannot exclude extra absorption using the X-ray data.  

For the June observation, the spectrum is, again, adequately fit with a
powerlaw model with absorption.  With $N_{\rm H} = 9.0 \times 10^{21}
\rm \, cm^{-2}$, the best fit photon index is $\Gamma = 1.77$ and the
allowed range at 90\% confidence is 1.52 to 2.02.   The spectrum in
June appears consistent with that measured in March.  The absorbed flux
is $(1.6 \pm 0.1) \times 10^{-13} \rm \, erg \, cm^{-2} \, s^{-1}$ in
the 0.3-8 keV band and the unabsorbed flux is $2.9 \times 10^{-13} \rm
\, erg \, cm^{-2} \, s^{-1}$ in the same band.

Since there is no evidence of spectral variability, we fitted the two
data sets simultaneously to obtain better constraints on the fit
parameters.  With $N_{\rm H} = 9.0 \times 10^{21} \rm \, cm^{-2}$, the
best fit photon index is $\Gamma = 1.77$ and the allowed range at 90\%
confidence is 1.64 to 1.90. Allowing the absorption column density to
vary leads to $N_{\rm H} = (9.9^{+2.9}_{-2.2}) \times 10^{21} \rm \,
cm^{-2}$  and $\Gamma = 1.87 \pm 0.31$ at 90\% confidence ($\Delta
\chi^2 = 4.6$ for two parameters).

In addition to the non-thermal powerlaw model, we also fit the combined
data with the mekal model which is appropriate for thermal emission
from hot diffuse gas.  With $N_{\rm H}$ fixed to the Galactic value, an
adequate fit was obtained with $kT = 5.6^{+2.2}_{-1.1} \rm \, keV$.   A
simple thermal bremsstrahlung model gives a similar temperature range,
$kT = 6.1^{+2.5}_{-1.5} \rm \, keV$.  No prominent line emission is
observed.  However, the limits on the equivalent width are not strongly
constraining (varying from 100~eV to 1.3~keV at 90\% confidence
($\Delta \chi^2 = 2.7$) for a narrow line with energy in the range
3--7~keV).

% \epsscale{1.5} for preprint2 1.0 for preprint

\begin{figure*}[t] \centerline{\epsscale{1.5} \plotone{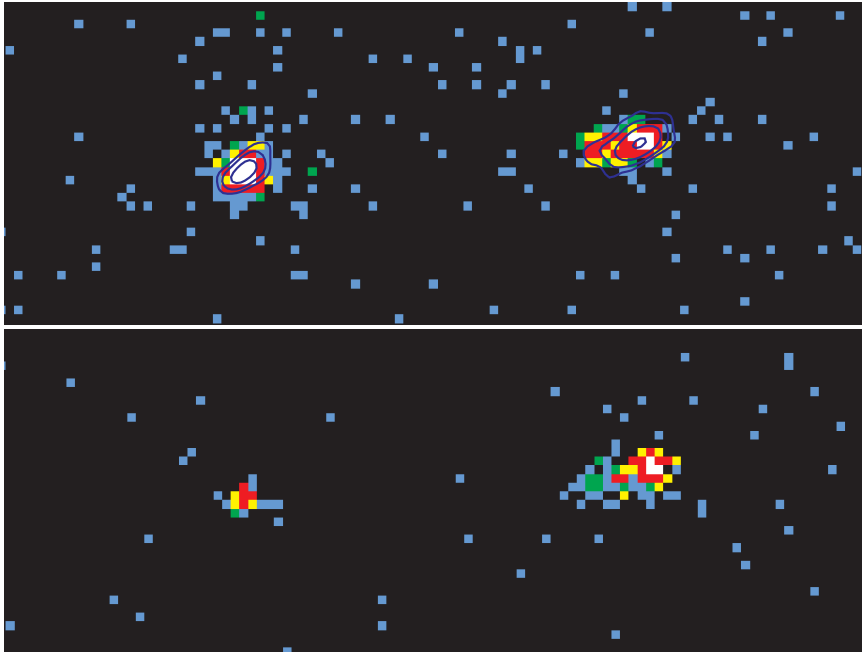}}
\caption{X-ray images of XTE J1550--564 (on the left = east) and the
western jet (on the right = west).  The upper panel is from the 11
March 2002 observation and the lower panel is from the 19 June 2002
observation. The X-ray data are for the 0.3--7~keV band and the color
scale is the same as in Fig.~\ref{ximage}.  The upper panel has with
radio contours (dark blue curves) from an observation on 29 January
2002 superimposed. The contour levels are 0.2, 0.4, 0.8, and 1.6~mJy. 
\label{xr}} \end{figure*}

\begin{figure}[t] \centerline{\epsscale{0.7} \plotone{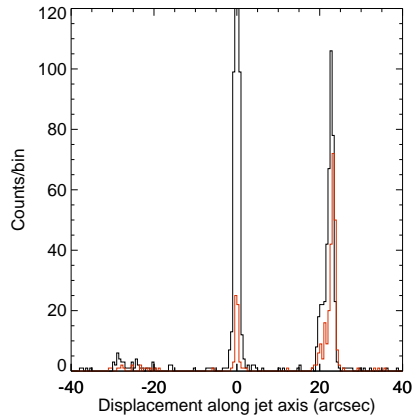}}
\caption{Distribution of X-ray counts along the jet axis.  The black
line is for the March observation and the red line is for the June
observation.  The bin size is $0.5\arcsec$. \label{profpar}}  
\end{figure}

\begin{figure}[t] \centerline{\epsscale{0.7} \plotone{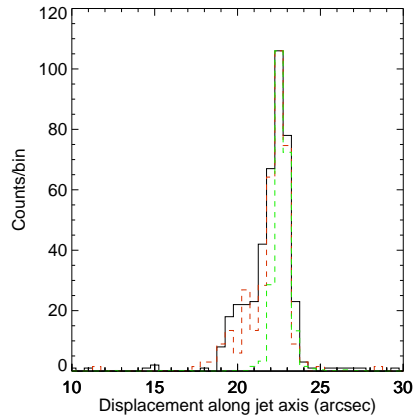}}
\caption{Distribution of X-ray counts in the western jet along the jet
axis.  The black (solid) line is for the March observation, the red
(dashed) line is for the June observation, and the green (dashed) line
is the profile of XTE J1550--564 from March.  The latter two profiles
have been shifted and rescaled to match the peak of the first.  The bin
size is $0.5\arcsec$. \label{profeast}}   \end{figure}

\begin{figure}[t] \centerline{\epsscale{0.7} \plotone{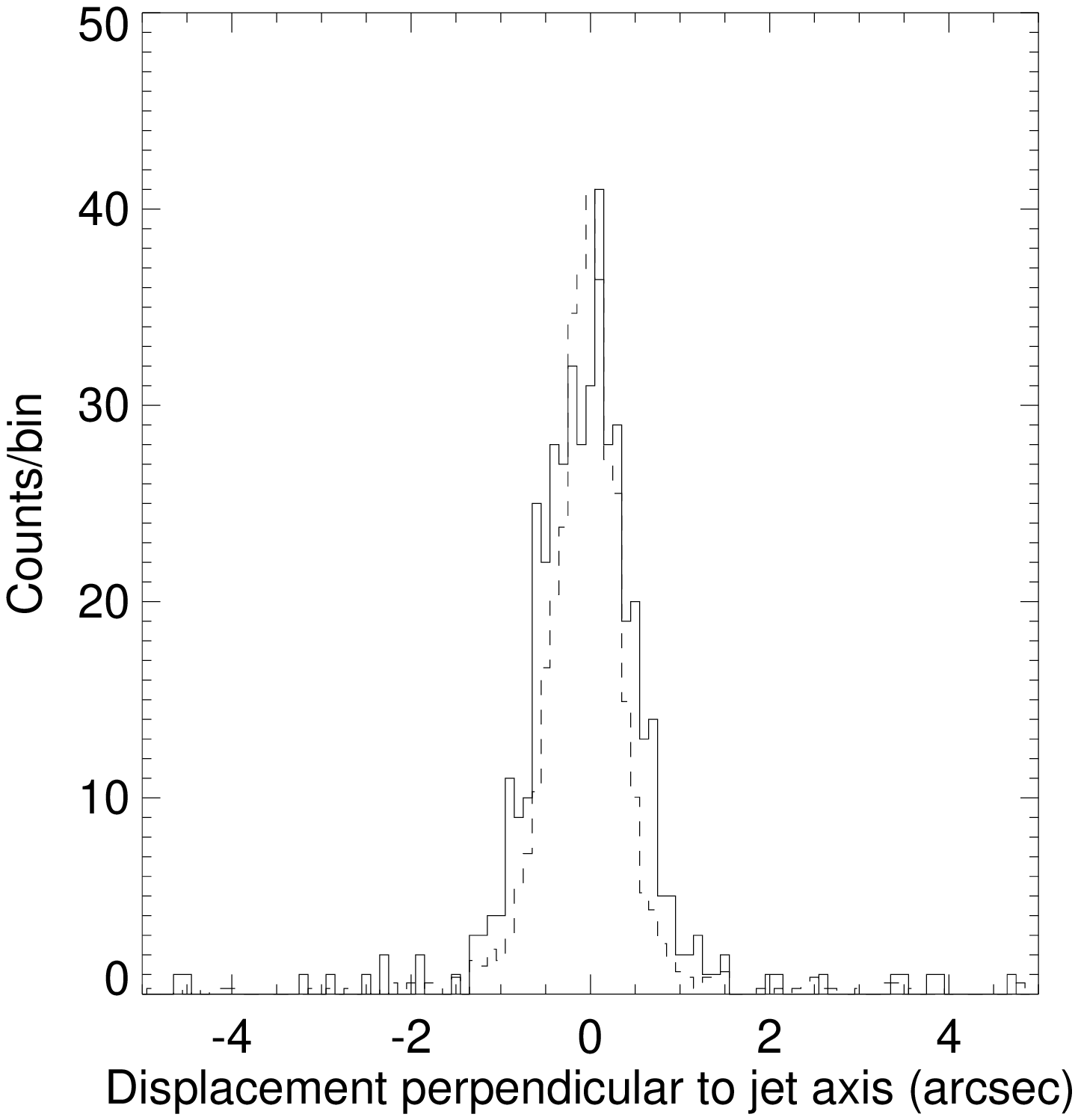}}
\caption{Distribution of X-ray counts perpendicular to the jet axis
from the March 2002 observation.  The solid line is for the western
jet.  The dashed line is the profile of XTE J1550--564 rescaled to
match the peak of emission in the western jet. The bin size is
$0.1\arcsec$. \label{profperp}}   \end{figure}

\subsection{Morphology and Motion}

The X-ray counterpart of the western jet appears extended. 
Fig.~\ref{xr} shows the X-ray images from both Chandra observations
with radio contours obtained from observations made on 29 January 2002
using the Australian Telescope Compact Array (ATCA) \citep{corbel_jet}
superimposed on the 11 March 2002 Chandra data. The ATCA image of the
compact source XTE J1550--564 appears extended along a NW-SE axis due
to partial synthesis caused by the limited parallactic angle coverage
with the linear ATCA array.  However, in addition to the extent caused
by partial synthesis, there is a true, physical extent to the western
radio jet.  The peak of the jet emission is towards the west (away from
XTE J1550--564) and there is lower intensity emission extending to the
east, back towards XTE J1550--564.  The morphology of the X-ray source
closely matches that of the radio source.

We define the jet axis using the positions of the X-ray sources
corresponding to XTE J1550--564 and the western jet, as reported in
Table~\ref{tablesrc}.  The resulting position angle is $-85.9\arcdeg
\pm 1.3\arcdeg$ in good agreement with the position angle of
$-86.1\arcdeg \pm 0.8\arcdeg$ reported for the superluminal jets (D.\
Hannikainen, private communication) and with the position angle of
$-85.8\arcdeg \pm 1.0\arcdeg$ reported for the radio western jet
\citep{corbel_jet}. To determine if the jet tails lie along the jet
axis, we calculated the average azimuthal angle for events between
$2\arcsec$ and $6\arcsec$ from the western jet position and within $\pm
40 \arcdeg$ in azimuth of a vector pointing back to XTE J1550--564. The
average azimuthal angle is $2.2\arcdeg \pm 1.9\arcdeg$ for the March
data and $7.4\arcdeg \pm 3.2\arcdeg$ for the June data.  The tail of
the western jet appears aligned with the jet axis in the March data,
but may be slightly skewed to the South in the June data.  However, 
the skew is only marginally statistically significant and may be
affected by uncertainty in the true position of the jet peak, discussed
below.

To study the morphology and motion of the X-ray jet, we decomposed the
image along axes parallel and perpendicular to  the jet axis (as
defined above).  We calculated the displacement of each X-ray event
parallel and perpendicular to this axis.  Fig.~\ref{profpar} shows the
profile of X-ray counts along the jet axis from the two observations. 
All photons with energies in the range 0.3--7~keV and within $2\arcsec$
of the jet axis in the perpendicular direction are included.  In the
March data, both jets are clearly present.  In the June data, there is
no strong source at the location of the eastern jet, but there may be
some diffuse emission above the background level.  The western jet
appears to have moved away from XTE J1550--564.

For the March observation, the peak of the X-ray emission of the
western jet is displaced by $0.6\arcsec$ (along the jet axis and away
from XTE J1550--564) from the position reported in Table~\ref{tablesrc}
which is the centroid over the region containing 50\% of the encircled
energy \citep{freeman02}.  Hence, caution is warranted when comparing
the jet position to positions found in other observations.

To measure the relative position of the western jet in the March and
June observations, we used a Kolmogorov-Smirnov (KS) test to permit
comparison of unbinned position data.  We used the position along the
jet axis for each event calculated above.  We added an offset, in the
range $-1\arcsec$ to $+1\arcsec$, to the June event positions and then
compared the offset positions with the March event positions within
$\pm 8\arcsec$ of the western jet peak.  The best match occurred for an
offset of $0.52\arcsec$.  Given this offset, the KS test gives a 61\%
probability that the two samples are drawn from the same parent
distribution.  To evaluate the uncertainty in the offset, we integrated
the KS-test probability distribution as a function of offset and found
the offsets corresponding to 5\% and 95\% of the full integral (i.e.\
the 90\% confidence interval).  These are $0.39\arcsec$ and
$0.60\arcsec$.  The hypothesis that the two distributions are the same
with zero offset is rejected at the level of $5 \times 10^{-13}$. The
western jet moved by $0.52\arcsec \pm 0.13\arcsec$ between the March
and June observations.

The morphology of the western jet does not appear to have changed, even
though it moved.  Fig.~\ref{profeast} shows the western jet profiles
along the jet axis from the March and June observations, with the June
data shifted by $0.52\arcsec$ and scaled so that the peaks match.  As
noted above, a KS test is consistent with the two profiles being the
same.  

The figure also shows the data for XTE J1550--564 itself from the March
observation, rescaled and displaced to match the peak at western jet. 
From its time variability, we know that the X-ray emission from XTE
J1550--564 must arise from what is effectively a point source for the
angular resolution of Chandra.  Hence, we use XTE J1550--564 as a
calibration of the point spread function for the observation to
determine the spatial extent of the emission from the western jet. 
This procedure accounts for any aspect jitter during the observation
and also for any source extent induced by scattering of the X-rays in
the interstellar medium between us and the source.  The spectrum of XTE
J1550--564 during the March observation is quite similar to that found
for the western jet, so effects due to the energy dependence of the
Chandra point spread function should be negligible.  Both sources are
sufficiently close to the aimpoint so that the degradation of the
Chandra point spread function off axis should also be negligible.

Comparison of the rescaled profile of XTE J1550--564 with the profile
of the X-ray emission from the western jet in the March observation
shows that the western jet is extended.  The leading edge of the X-ray
source (the edge away from XTE J1550--564) is approximately consistent
with the profile of a point source.  The trailing edge is clearly
extended.  The full width of the source at half maximum is $1.2\arcsec$
and the full width at 10\% of maximum intensity is $5\arcsec$.

Fig.~\ref{profperp} shows the distribution of X-ray counts
perpendicular to the displacement axis for the western jet and XTE
J1550--564 in the March data.  We used X-rays with energies in the
range 0.3--7~keV and with displacements along the jet axis of up to
$\pm 5\arcsec$ from the respective source position as reported in
Table~\ref{tablesrc}.  A Kolmogorov-Smirnov (KS) test shows that the
two distributions are inconsistent at the 97\% confidence level ($2.2
\sigma$).  The western jet may be slightly extended perpendicular to
the jet axis, but the evidence is weak.  Deconvolving, assuming that
the physical width adds in quadrature with the instrumental width, we
place an upper limit of $0.8\arcsec$ (FWHM) on the extent of the X-ray
western jet perpendicular to the jet axis. The perpendicular
distribution of the western jet found from the June data is consistent
with that in the March data.  The X-ray western jet is clearly much
more extended along the jet axis than perpendicular to it.

\subsection{Variability}

To investigate the variability of the X-ray western jet, we examined
archival Chandra data on XTE J1550--564 from June, August, and
September 2000.  Using a large region extending $\pm 4\arcsec$
perpendicular to the jet axis and covering the full extent of the
western jet beginning $11\arcsec$ from XTE J1550--564 and extending
$2\arcsec$ beyond its position in 2002, we find no evidence for X-ray
emission from the western jet in any of the archival observations with
upper limits ($2\sigma$) on the ACIS counting rate which are fractions
of 0.10-0.16 of the counting rate in our March 2002 data.  Combining
the August and September 2000 observations, in which the ACIS was not
partially blocked by a grating, we find 6 counts in a total exposure of
9557~s.  Given a mean of 6, the 90\% confidence upper bound from a
Poisson distribution is 9 counts leading to a 90\% confidence upper
bound on the count rate from the western jet of 0.00094~c/s.  This is
6\% of the rate in the March 2002 observation, corresponding to an
upper bound on the absorbed flux (assuming identical spectral
parameters) of $1.1 \times 10^{-14} \rm \, erg \, cm^{-2} \, s^{-1}$ in
the 0.3--8 keV band.  This analysis assumes that none of the counts are
due to background.  Performing a background subtraction would reduce
this limit.

During 2002, the flux decayed between March and June.  Using the two
flux measurements to calculate the $1/e$-folding time for an
exponential decay, we find $327 \pm 95 \rm \, days$.  However, two data
points are insufficient to determine the shape of light curve decay.

\begin{figure}[t] \centerline{\epsscale{0.7} \plotone{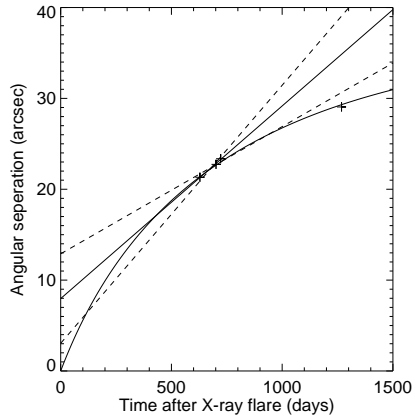}}
\caption{Position of the centroid of the X-ray eastern jet versus time.
The solid line is the fit to the proper motion in the 2000 data found
by \citet{tomsick_jet}.  The dashed line indicate the uncertainty in
the fit parameters.  The solid curve is a decelerating jet model fit
described in the text. \label{motion}}   \end{figure}

\section{Eastern Jet}

The new Chandra observations add one additional detection to the
results presented in \citet{tomsick_jet}.  The source contains far
fewer counts than the western jet and so we concentrate on the position
and flux of the source.

The proper motion of the eastern jet is shown in Fig.~\ref{motion}. The
figure includes the three data points from \citet{tomsick_jet} and our
new data point for 11 March 2002.  \citet{tomsick_jet} showed that the
apparent velocity of the eastern jet in 2000 was lower than the minimum
velocity allowed during the superluminal ejection
\citep{hannikainen01}, thus indicating that the eastern jet has
decelerated.  The March 2002 point is inconsistent with an
extrapolation of the velocity measured from the 2000 data and indicates
that this deceleration has continued.

We fit the proper motion data with a model in which the jet has a
deceleration proportional to its velocity relative to the X-ray binary
(which we assume to be roughly at rest relative to the interstellar
medium).  The velocity profile is then a decaying exponential and the
observed position evolution is modified by light travel delays.  Our
model has three fitted parameters: the initial jet speed divided by the
speed of light $\beta_0$, the time scale ($1/e$-folding time) for the
velocity decay $\tau$, and the jet angle relative to the line of sight
$\theta$.  The assumed source distance affects the fitted parameters. 
For a source distance of 4~kpc, we find an adequate fit (shown in
Fig.~\ref{motion}) with $\beta_0 = 0.94$, $\tau = 1030 \rm \, days$,
and  $\theta = 62\arcdeg$.  For source distances larger than 5.1~kpc,
the best fit initial speed is larger than the speed of light.  However,
we do not consider this a constraint on the distance as the model is
rather ad hoc.  For a distance of 3~kpc, the best fit is $\beta_0 =
0.82$, $\tau = 960 \rm \, days$, and $\theta = 50\arcdeg$.  Both of
these fits are consistent with the lower bound on the initial jet speed
from the VLBI observations \citep{hannikainen01}.

\begin{figure}[t] \centerline{\epsscale{0.7} \plotone{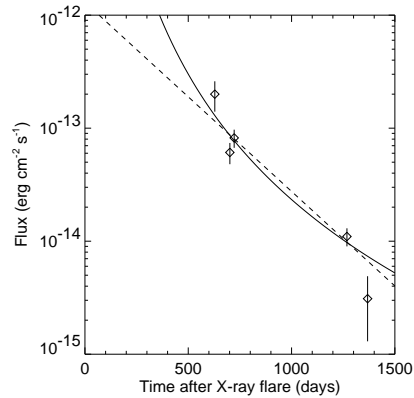}}
\caption{X-ray flux of the eastern jet versus time. The curves are the
powerlaw (solid) and exponential (dashed) decays described in the text.
\label{fluxv}}   \end{figure}

Fig.~\ref{fluxv} shows the time variation of the X-ray flux of the
eastern jet.  The first three points are from \citet{tomsick_jet}. For
consistency with that analysis, we found the fluxes for the March 2002
data using a circular extraction region of $4\arcsec$ radius centered
on the wavdetect position and with an annulus with an inner radius of
$5\arcsec$ and an outer radius of $18\arcsec$ for background
estimation.  We used a powerlaw model with photon index fixed to 1.8
and an absorption column density fixed to $9\times 10^{21} \rm \,
cm^{-2}$ and corrected for the degradation in the low energy quantum
efficiency of the ACIS.  The absorbed flux in the 0.3-8~keV band was
$(1.1 \pm 0.3) \times 10^{-14} \rm \, erg \, cm^{-2} \, s^{-1}$.  For
June, no source was detected by {\it wavdetect} at the location of the
western jet. We extracted counts from a circular region $4\arcsec$
radius centered at the position found in March.  This region is
sufficiently large to include most of the diffuse emission apparent in
Fig.~\ref{profpar}. With spectral parameters fixed as above, the
absorbed flux in the 0.3-8~keV band is $(3 \pm 2) \times 10^{-15} \rm
\, erg \, cm^{-2} \, s^{-1}$.  We interpret this number as an upper
limit on the jet flux in June.

We fit both an exponential decay and a powerlaw decay to the flux data.
The powerlaw decay provides a slightly better fit, but both fits are
acceptable.  The index of the powerlaw decay is $3.7 \pm 0.7$.  The
$1/e$-folding time of the exponential decay is $260 \pm 50 \rm \, days$.
The extrapolated flux at the origin of the jet for the exponential
decay would be $(1.3 \pm 0.5) \times 10^{-12} \rm \, erg \, cm^{-2} \,
s^{-1}$ in the 0.3-8~keV band.  This is a small fraction of the X-ray
fluxes measured during the 1998 outburst.  Hence, the available X-ray
data are not inconsistent with an exponential decline of flux since the
origin of the eastern X-ray jet.

\section{Discussion}

The discovery of extended radio and X-ray emission from XTE J1550--564
(Corbel et al.\ 2002b, Tomsick et al.\ 2002, and the results reported
here) represents the first detection of large scale relativistic jets
from a Galactic black hole candidate in both radio and X-rays.  These
large scale jets appear to arise from a relatively brief jet injection
episode \citep{corbel_jet} and, therefore, offer a unique opportunity
to study the large-scale evolution of an impulsive jet event.

In the following, we assume that both jets were created in a single
injection episode in September 1998.  The assumption is motivated by
the detection of superluminal jets \citep{hannikainen01} following an
extremely large X-ray flare in September 1998 and the absence of any
other X-ray flare of similar magnitude in continual X-ray monitoring
from 1996 to 2002 \citep{corbel_jet}.  The eastern jet appears to be
the approaching jet based on its larger current separation from XTE
J1550--564 \citep{corbel_jet}.  The distance to XTE J1550--564 is
constrained from optical observations to be in the range 2.8--7.6~kpc
with a favored value of 5.3~kpc \citep{orosz02}.  

We detected X-ray emission from the western jet on 11 March 2002 (MJD
52344) at a separation of $23\arcsec$ from the black hole candidate
implying a projected physical separation of $0.59 {\, \rm pc} (d/5.3 \,
{\rm kpc}) = 1.8 \times 10^{18} {\, \rm cm} (d/5.3 \, {\rm kpc})$ and a
mean projected speed of $18.1 \pm 0.4 \rm \, mas/day$ or $0.55 c (d/5.3
\, {\rm kpc})$.  From the  motion of the jet between 11 March and 19
June, we calculate an average speed at that epoch of $5.2 \pm 1.3 \rm
\, mas/day$ or $(0.16 \pm 0.04) c (d/5.3 \, {\rm kpc})$.  This is
significantly less than the average speed from 1998 to 2002 and
indicates that the jet has decelerated.  Our detection of motion in the
western jet argues against it having reached a termination.   The jet
appear to have a relativistic bulk velocity.

The angular size of the X-ray emission perpendicular to the jet axis in
the western jet is quite small, $< 0.8\arcsec$ (FWHM), placing an upper
bound on the jet half opening angle of $1.0\arcdeg$.  The opening angle
is small, similar to the half opening angles of Fanaroff-Riley class II
(FR-II; Fanaroff \& Riley 1974) sources  which are less than $3\arcdeg$
\citep{muxlow91}.  The angular size of the western X-ray jet
perpendicular to the jet axis limits the expansion velocity to less
than $0.01 c (d/5.3 \rm kpc)$.  At the same epoch, the eastern jet has
a projected physical separation of $0.75 {\, \rm pc} (d/5.3 \rm kpc)$.

For comparison with the properties of the western jet, we review the
properties of the eastern jet at the same angular separation.  From the
linear fit in \citet{tomsick_jet}, we find that the eastern jet passed
through a separation of $23\arcsec$ on MJD $51785 \pm 16$. 
Fortuitously, this is consistent with the times of the 21 August 2000,
and 11 September 2000 Chandra observations (the separations from both
of those observations are consistent with $23\arcsec$).  As discussed
in \citet{tomsick_jet}, the high proper motion of the eastern jet when
at a separation of $23\arcsec$ argues strongly against it having
reached a termination at that point.  The continued motion of the
eastern jet to larger separations demonstrated here reinforces this
argument.  At a separation of $23\arcsec$, the apparent speed of the
eastern jet was $21.2 \pm 7.2 \rm \, mas/day$ and the absorbed flux was
$(7.2 \pm 1.0) \times 10^{-14} \rm \, erg \, cm^{-2} \, s^{-1}$ in the
0.3-8 keV band.  The western (receding) jet flux at the same separation
was a factor of $2.6 \pm 0.4$ higher than the eastern (approaching) jet
flux.

The X-ray data provide measurements of the local apparent speed at
equal angular separations from the origin for the approaching and
receding jets.  If the jets are symmetric, in terms of their velocity
profiles, then it would be possible to use relativistic kinematics
\citep{mirabel99} to constrain the jet true speed $\beta$, inclination
to the line of sight $\theta$, and distance $d$.  We find that $\beta
\cos \theta = 0.61 \pm 0.13$ and that $d = (12.6 \pm 4.4) \tan \theta
\rm \, kpc$ under the assumption that the jet propagation is symmetric.
This would imply that the inclination angle $\theta \le 61\arcdeg$ and
that the jet speed $\beta \ge 0.48$ (allowing for the uncertainty in
$\beta \cos \theta$).

However, the fact that the western (receding) jet is brighter than the
eastern (approaching) jet argues against symmetric jet propagation. 
For symmetric jet propagation, the ratio of observed flux densities
measured at equal separations from the core for a twin pair of
optically-thin isotropically emitting jets is

\begin{equation} 
\frac{S_a}{S_r} = \left( \frac{1 + \beta \cos \theta}{1 - \beta \cos
\theta} \right) ^ {k - \alpha}
\end{equation}

\noindent where $S_a$ is the approaching flux density, $S_r$ is the
receding flux density, $\beta$ is the jet speed divided by the speed of
light, $\theta$ is jet inclination angle, $\alpha$ is the spectral
index of the emission defined so $S_{\nu} \propto \nu^\alpha$, and $k$
is a parameter which is 2 for a continuous jet and 3 for discrete
condensations \citep{mirabel99}.  If the exponent $k-\alpha > 0$, then
the approaching jet must be always be brighter than the receding jet
measured at the same angular separation from the core.   Given the
photon index quoted above (allowing the absorption to vary) and
allowing a continuous jet, the minimum allowed value for the exponent
is 2.6 which is well above zero. Hence, the observed brightness ratio
$S_a/S_r = 0.38 \pm 0.05$ is inconsistent with symmetric jet
propagation.  However, because the X-ray jet detections occurred well
after the initial jet ejection, we cannot independently constrain the
jet ejection, which may have been symmetric.

Large intrinsic asymmetries have been observed in the radio for the
jets from the microquasar GRO J1655--40 \citep{hjellming95}. Which of
the two jets was brighter differed from day to day, while the
kinematics of the jet propagation appeared symmetric.  Asymmetries of
similar magnitude could explain the ratio of X-ray fluxes from the jets
of XTE J1550--564 if the jet inclination angle is high.  In this case,
the X-ray emissivity of the western jet would need to be low early on,
to be consistent with the non-detection in 2000, and then increase
sharply at later times, to be consistent with the flux measured in
2002.  Such an X-ray emissivity profile could be produced by, e.g.,
internal shocks produced by a faster plasmon overtaking a slower one. 
This would require re-acceleration of particles in the jet long after
its initial ejection.

Alternatively, the emission from the jets could be produced by shocks
arising from interactions of the jets with the interstellar medium
(ISM).  The jets appear to be moving with speeds much greater than the
sound speed of the ISM.  In this case, the supersonic motion of the jet
should produce a shock wave.  The shock will be strongest at the head
of the head and weaken around the sides of the jet \citep{deyoung02}.
This morphology matches that observed for the eastern jet.  In this
case, the asymmetry between the two jets may reflect non-uniformity in
the ISM.  If the western jet is propagating into a denser medium, it
would be brighter and would have decelerated more than the eastern jet.
Interaction of a relativistic plasmon with the ISM appears to
consistently describe all of the available data on the large scale jets
of XTE J1550--564.

Continued observation of the jets of XTE J1550--564 offers an exciting
opportunity to study the dynamics of relativistic jets on time scales
inaccessible for AGN jets.  New, deep observations of XTE J1550--564
will allow us to study the deceleration and evolution of the
morphology, flux, and spectrum of the western jet.  This may help
determine whether the observed jets are due to the internal
interactions of relativistic plasmons or due to external interactions
with the ISM.  A deep observation may also permit a new detection of
the eastern jet which would further constraint its deceleration and
decay.  Finally, continued monitoring may eventually show the
termination of the jets.

\acknowledgments

We thank Harvey Tananbaum for granting the DDT observations, Joy
Nichols for exceptionally rapid processing of the data, and the CXC
team for successfully executing the observation.  PK thanks Heino
Falcke, Dan Harris, and David De Young for useful discussions and
acknowledges partial support from NASA grant NAG5-7405 and Chandra
grant number G01-2034X.  RW was supported by NASA through Chandra
Postdoctoral Fellowship grant number PF9-10010 awarded by CXC, which is
operated by SAO for NASA under contract NAS8-39073.

\end{document}